\documentclass[12pt]{article}
\usepackage{epsfig}

\begin{document}
\begin{titlepage}

\centerline{\large\bf Center vortex model for the infrared sector of}
\vspace{0.2cm}
\centerline{\large\bf SU(3) Yang-Mills theory -- baryonic potential}

\bigskip
\centerline{M.~Engelhardt\footnote{\tt email:\ engel@nmsu.edu} }
\vspace{0.2 true cm}
\centerline{\em Physics Department, New Mexico State University}
\centerline{\em Las Cruces, NM 88003, USA}

\begin{abstract}
The baryonic potential in the framework of the $SU(3)$ random vortex
world-surface model is evaluated for a variety of static color source
geometries. For comparison, carefully taking into consideration the
string tension anisotropy engendered by the hypercubic lattice description,
also the $\Delta $ and Y law predictions for the baryonic potential are
given. Only the Y law predictions are consistent with the baryonic
potentials measured.
\end{abstract}

\vspace{1cm}

{\footnotesize PACS: 12.38.Aw, 12.38.Mh, 12.40.-y}

{\footnotesize Keywords: Center vortices, infrared effective theory,
confinement, baryon potential}

\end{titlepage}

\section{Introduction}
The random vortex world-surface model of the Yang-Mills vacuum 
\cite{vormod,topol,csb,su3} is designed
to describe the central phenomena induced by the strong interaction in
the infrared regime: Confinement, spontaneous breaking of chiral symmetry,
and the axial $U_A (1)$ anomaly. The motivation for the model is drawn from
lattice Yang-Mills studies 
\cite{deb97,giedt,fabdo3,greensrev,bertop,temp,tlang,forc1,forc2,pepe}
which suggest the relevance of center vortices for infrared strong
interaction physics. For the case of $SU(2)$ color, the model has been shown
to correctly predict (in the sense of agreement with corresponding lattice
Yang-Mills results) a number of observables characterizing the aforementioned
phenomena.

In a recent initial study of the generalization to $SU(3)$ color \cite{su3},
the random vortex world-surface model was again found to reproduce the
confinement properties of Yang-Mills theory. Both a low-temperature
confining phase as well as a high-temperature deconfined phase are
generated, separated by a weakly first order phase transition. The
confinement properties are intimately related to the percolation properties
of the vortices.

\begin{figure}
\centerline{\epsfig{file=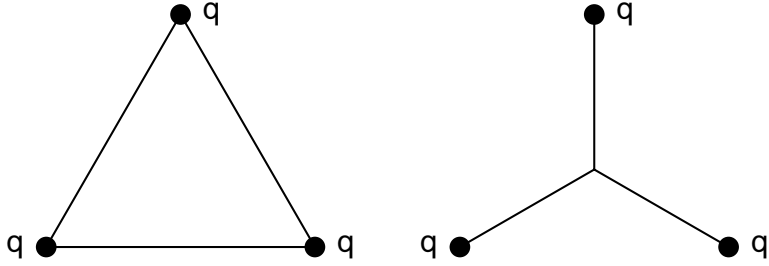,width=11cm} }
\caption{\footnotesize
Based on the notion that three $SU(3)$ color sources in the
Yang-Mills vacuum are bound by (linear) potentials associated with distances
characterizing the spatial geometry of the sources, two distinct model bond 
topologies can be envisaged: The $\Delta $ law, where sources are connected
pairwise (left), and the Y law, where bonds emanating from the sources
meet at an (optimally placed) central junction (right). While the string
tension in the Y law must be the same as in a mesonic configuration
(where also a single bond emanates from each color source), the $\Delta $
model is assumed to be associated with half that string tension. This
is based both on the short-distance perturbative limit and on the limit
where two color sources approach each other, which should yield the usual
mesonic string tension. Since in this limit, two of the bonds making up
the $\Delta $ configuration merge, each of them is taken to contribute half 
of the mesonic string tension (this linear superposition of
bonds constitutes an auxiliary model assumption which is
not readily justified within the underlying nonabelian Yang-Mills theory,
in which chromoelectric flux tubes interact nontrivially when approaching
each other).}
\label{fig1}
\end{figure}

Apart from the order of the deconfining transition, a further aspect in
which the $SU(3)$ theory differs substantially from the $SU(2)$ case is the
long-range baryonic potential, which was not studied in \cite{su3}. This
constitutes the subject of the present work. To arrive at definite statements
concerning the baryonic potential, two issues must be taken into account.
One issue is the breaking of spatial rotational symmetry in the present
hypercubic lattice formulation of the model; after all, to measure baryonic
potentials, it is necessary to evaluate Wilson loops which are not located
within two-dimensional lattice planes. On the other hand, the Wilson loop
areas spanned by baryonic configurations are substantially larger than the
ones encountered in simple mesonic Wilson loop calculations; as a result,
it is necessary to employ appropriate numerical noise reduction techniques
to extract a meaningful signal for the baryonic potential. Having dealt with
these two issues, the data collected in this work unequivocally point to
a Y law for the baryonic potential as opposed to a $\Delta $ law,
cf.~Fig.~\ref{fig1}.
It should be noted that the form of the baryonic potential generated by
vortex models has been the subject of recent debate \cite{corndlaw,cornylaw};
on the other hand, also within full $SU(3)$ lattice Yang-Mills theory, only
quite recently sufficiently large separations have been probed to clearly
detect a Y law \cite{takahashi,phil01,phil02,philoli}. In the full theory,
short-distance perturbative $\Delta $ behavior presumably masks the
long-range Y law such that a crossover between the two dependences is
observed at intermediate distances \cite{phil02} and the Y law emerges
only at rather large separations. Aside from the vortex model, also
other models of the strong interaction vacuum yield a Y law for the baryonic
potential, e.g.~the dual superconductor picture \cite{cherkom,ichie} and
the stochastic vacuum \cite{stochv}.

\section{Model description}
\label{moddesc}
The random vortex world-surface model is based on the notion that the
Yang-Mills vacuum is populated by random closed lines of quantized
chromomagnetic flux (vortices), described by an ensemble of closed
random world-surfaces in four-dimensional (Euclidean) space-time.
In the present implementation of the model, these random surfaces are
generated on a hypercubic lattice, composed of elementary squares on
that lattice. The action governing the ensemble is related to the surface
curvature: If two elementary squares which are part of a
vortex surface share a lattice link but do not lie in the same plane, 
this costs an action increment $c$. Formally, this can be written as a
sum over lattice links,
\begin{eqnarray}
S[q] \! \! \! \! &=& \! \! \! \!
c \sum_x\sum_\mu \left[ \sum_{\nu < \lambda \atop \nu \neq \mu,
\lambda\neq \mu} \left( | q_{\mu\nu}(x) \, q_{\mu\lambda}(x) |
 + | q_{\mu\nu}(x) \, q_{\mu\lambda}(x-e_\lambda) | 
\right. \right. \label{curvature} \\
& & \ \ \ \ \ \ \ \ \ \ \ \ \ \ \ \ \ \
+ \left. | q_{\mu\nu}(x-e_\nu) \, q_{\mu\lambda}(x) |
 + | q_{\mu\nu}(x-e_\nu) \, q_{\mu\lambda}(x-e_\lambda) |
\right)\Bigg] \nonumber \\
&=& \! \! \! \! \frac{c}{2} \sum_{x}\sum_\mu \left[ \left[
\sum_{\nu\neq\mu} \left( | q_{\mu\nu}(x) | +
| q_{\mu\nu}(x-e_\nu) | \right) \right]^2 \! \! \! - \! \!
\sum_{\nu\neq\mu} \! \! \left[ | q_{\mu\nu}(x) | +
| q_{\mu\nu}(x-e_\nu) | \right]^2 \ \right] \ , \nonumber
\end{eqnarray}
where $q_{\mu \nu } (x)$ describes the chromomagnetic flux associated with
the elementary square extending from the site $x$ into the positive $\mu $
and $\nu $ directions. In the $SU(3)$ model, the variables $q_{\mu \nu } (x)$
can take three values, $q_{\mu \nu } (x) \in \{ -1,0,1 \} $; the value $0$
indicates that the elementary square is not part of a vortex surface, whereas
the values $\pm 1$ indicate that it is. The latter two values distinguish
the two possible types of quantized flux carried by $SU(3)$ vortices. These
will be characterized in detail in the next section by their effect on
Wilson loops circumscribing them. Note that $S$ is symmetric with respect
to the two possible types of vortex flux. Note furthermore that, formally,
there are two variables connected with the elementary square extending from
the site $x$ into the positive $\mu $ and $\nu $ directions, namely
$q_{\mu \nu } (x)$ and $q_{\nu \mu } (x)$; these correspond to the two
opposite orientations, i.e.~senses of curl, one may ascribe to a surface.
However, since both variables are meant to describe the same flux (which,
apart from its location in space-time, also carries a definite orientation),
they are related by\footnote{Note that, in light of the description above,
this (correctly) implies that a vortex carrying one type of flux is
equivalent to an oppositely oriented vortex carrying the other type of
flux for the present purposes (namely, for evaluating Wilson loops).
Indeed, these two cases only differ by a flux $2\pi $, i.e.~a Dirac string,
which is unobservable in Wilson loops. On the other hand, when considering
topological properties \cite{cont,topol}, the two cases do have to be
distinguished and one cannot trade off geometrical orientation of a vortex
in space-time against the magnitude of flux it carries.}
$q_{\mu \nu } (x) = -q_{\nu \mu } (x)$.

It should also be emphasized that the vortex ensemble is generated
subject to the constraint of continuity of flux (modulo $2\pi $,
i.e.~modulo Dirac strings \cite{topol,csb,su3}). For one, this forces
vortex surfaces to be closed, as already mentioned above; on the other hand,
the existence of two types of quantized flux in the $SU(3)$ model nevertheless
allows for vortex branchings, where a vortex of one type splits into two
vortices of the other type while maintaining continuity of flux\footnote{This
does not happen in the $SU(2)$ model, which only has one type of flux,
corresponding to the existence of only one nontrivial element in the center
of the $SU(2)$ group.}. As described in detail in \cite{su3}, continuity
of flux in practice is guaranteed during the generation of the vortex
world-surface ensemble by performing updates simultaneously on the six
squares making up the surface of an elementary three-dimensional cube in the
lattice. This is done in a way which corresponds to superimposing the
(continuous) flux of a vortex of the shape of the elementary cube surface
onto the flux previously present.

The flux world-surfaces defined in this manner are formally
infinitely thin; however, they are meant to describe physical vortex
structures present in the Yang-Mills vacuum which possess a finite
transverse thickness, akin to the structures making up the Copenhagen
vacuum \cite{spag}. The model surfaces represent the geometrical centers
of such physical thick vortices. The thickness enters the model description
through the spacing of the lattice on which the vortex surfaces are generated.
For instance, two parallel physical vortices can only be meaningfully
distinguished from one another when they are further apart than a minimal
distance. In the model, this distance is encoded in the lattice spacing;
at shorter distances, the thick structures cannot be distinguished and
thus should be represented instead by a single model world-surface of
combined flux. For a thorough discussion of the role of the lattice
spacing in the random vortex world-surface model, cf.~\cite{vormod,su3}.

Given this definite physical meaning of the lattice spacing, it follows
that it is a fixed finite quantity in the model\footnote{This should also
be expected from the point of view that the random vortex world-surface
model is an infrared effective model, which is only defined up to a
fixed ultraviolet cutoff.}, which should be contrasted with conventional
lattice gauge theory, where one is ultimately interested in the limit of
vanishing lattice spacing. This has consequences as far as the breaking of
spatial rotational symmetry by the underlying hypercubic lattice structure
is concerned. Whereas in lattice gauge theory, spatial rotational symmetry
is restored in the course of taking the continuum limit \cite{kogut}, in
the random vortex world-surface model the breaking of rotational symmetry
engendered by the hypercubic lattice description and manifest in the action
(\ref{curvature}) is a real model property which will be quantified in
detail in section \ref{rotsb}. Roughly, string tensions along diagonal
directions in the lattice are enhanced compared with the one measured
along lattice axes to the extent that the most extreme string tension
values obtained are separated from their mean by 13~\%. This variation
must be taken into account when comparing baryonic potentials with
$\Delta $ or Y law expectations.

It should be noted that, while the random vortex world-surface model is
intended to be more than just a qualitative model of strong interaction
physics, on a quantitative level it is generally not expected to furnish
a description of medium- to long-range observables better than to within
of the order of 10~\%. After all, it encodes a truncated dynamics in
which phenomena such as the Coulomb potential between color sources are
missing and effects at distances shorter than about 0.4 fm cannot be resolved.
In light of this, the consequences of the hypercubic realization of the
random vortex surfaces, which include the breaking of spatial rotational
symmetry discussed above, represent just one particular model distortion of
reality among others. No more (or less) significance needs to be attached
to this specific distortion than to others on a similar quantitative level,
such as ambiguities in measurements of the topological susceptibility
\cite{topol} and of the quenched chiral condensate \cite{csb}. Note
also that the breaking of rotational symmetry is only introduced by
the particular (technically convenient) hypercubic lattice implementation
of the model; a manifestly symmetric random vortex world-surface model
could e.g.~be formulated using randomly triangulated surfaces in continuous
four-dimensional space-time.

\section{Observables}
\label{obsdesc}
Physically, the random vortex surfaces represent quantized chromomagnetic
flux. This means that they contribute in a characteristic way to Wilson
loops; if one chooses an area spanning a given Wilson loop\footnote{The
choice of area is immaterial due to the continuity of flux.}, then for each
time a vortex world-surface pierces\footnote{Note that Wilson loops are
defined on a lattice dual to the one on which the vortices are defined,
i.e.~on a lattice shifted by the vector $(a/2,a/2,a/2,a/2)$, where $a$
denotes the lattice spacing. Thus, the notion of a vortex piercing a
Wilson loop area is unambiguous.} that area, the Wilson loop acquires
a phase factor corresponding to the center of the gauge group. For the
case of $SU(3)$ color, there are two possible phase factors, associated
with the two possible types of quantized vortex flux,
namely $\exp (\pm i 2\pi /3)$. To be precise, consider an elementary
Wilson loop (plaquette) extending from $y$ into the positive $\kappa $
and $\lambda $ directions, with the integration oriented such that one
starts at $y$, integrates first into the positive $\kappa $ direction,
and then onwards around the plaquette. Denote this by
$U_{\kappa \lambda } (y)$. The plaquette $U_{\kappa \lambda } (y)$ is
pierced precisely by the dual lattice elementary square $q_{\mu \nu } (x)$,
where the indices $\kappa , \lambda , \mu , \nu $ span all four space-time
dimensions and $x=y+(\vec{e}_{\kappa } + \vec{e}_{\lambda } -
\vec{e}_{\mu } - \vec{e}_{\nu } ) a/2$, with $a$ denoting the lattice
spacing. $U_{\kappa \lambda } (y)$ can thus be given exclusively in
terms of the corresponding $q_{\mu \nu } (x)$, namely
\begin{equation}
U_{\kappa \lambda } (y) = \exp \left( i \pi / 3 \cdot
\epsilon_{\kappa \lambda \mu \nu } \,q_{\mu \nu }(x)\right)
\label{plaq}
\end{equation}
(with the usual Euclidean summation convention over Greek indices).

To evaluate an arbitrary Wilson loop, it is sufficient \cite{su3} to find
a tiling of that Wilson loop by a set of plaquettes and multiply the values
those plaquettes take in the given vortex configuration, as specified by 
(\ref{plaq}).

Consider now a baryonic Wilson loop,
\begin{equation}
W = \frac{1}{6} \epsilon_{abc} \epsilon_{a^{\prime } b^{\prime } c^{\prime } }
\Gamma_{1}^{aa^{\prime } } \Gamma_{2}^{bb^{\prime } }
\Gamma_{3}^{cc^{\prime } }
\label{bardef}
\end{equation}
where $\Gamma_{i} $ denotes the path-ordered exponential (Wilson) line
integral along the path taken by the $i$-th color source. All three
$\Gamma_{i} $ start at the same initial space-time point and end at the
same final space-time point. Due to the special properties of center
vortex configurations, the baryonic Wilson loop in such configurations can
be straightforwardly decomposed into three standard mesonic Wilson loops.
On the one hand, any center vortex configuration can be written in
terms of gauge fields defined purely within the (Abelian) Cartan
subgroup of the underlying gauge group \cite{cont}. Under these
circumstances, the $\Gamma_{i} $ in (\ref{bardef}) are diagonal, and
by multiplying (\ref{bardef}) with the (unit) determinant of another
color matrix $\Gamma_{0} $ given by an additional exponential line
integral connecting the end point of the $\Gamma_{i} $ with their
initial point,
\begin{equation}
1=\Gamma_{0}^{11} \Gamma_{0}^{22} \Gamma_{0}^{33}
\end{equation}
one can rewrite the baryonic Wilson loop as
\begin{equation}
W = \frac{1}{6} \sum_{a,b,c} \epsilon^{2}_{abc}
(\Gamma_{1}^{aa} \Gamma_{0}^{aa} )
(\Gamma_{2}^{bb} \Gamma_{0}^{bb} )
(\Gamma_{3}^{cc} \Gamma_{0}^{cc} )
\end{equation}
On the other hand, since in a center vortex configuration, the exponential
integral along a path with coinciding initial and final points is
proportional to the unit matrix (this is the defining property of center
flux), this indeed is equal to a product of three standard mesonic Wilson
loops,
\begin{equation}
W = \left( \frac{1}{3} \, \mbox{Tr} \, \Gamma_{1} \Gamma_{0} \right)
\left( \frac{1}{3} \, \mbox{Tr} \, \Gamma_{2} \Gamma_{0} \right)
\left( \frac{1}{3} \, \mbox{Tr} \, \Gamma_{3} \Gamma_{0} \right)
\end{equation}

In practice, in the present work, the baryonic correlator of three
Polyakov loops was evaluated. This is of course nothing
but a special case of the baryonic Wilson loop discussed
above, with paths $\Gamma_{i} $ starting from an initial
spatial position $x_0 $ at time $t=0$, running to
spatial positions $x_i $ while at $t=0$, remaining there
as time runs through the complete extension of the lattice
to the final time $t=\beta $ (which is identified with $t=0$,
with periodic boundary conditions on the gauge fields),
and returning to $x_0 $ along the same path as they took
at $t=0$. Due to the Abelian nature of the vortex gauge fields, all the
line integrals at $t=0$ and $t=\beta $ cancel and one is left with
a baryonic Polyakov loop correlator. In accordance with 
the discussion above, this is furthermore equal
to the expectation value of a product of three standard
mesonic Polyakov loop pairs such as the ones used to evaluate standard
mesonic Polyakov loop correlators. The baryonic Polyakov loop correlator
observable has the advantage that boundary effects resulting from the
propagation of the color sources from $x_0 $ to $x_i $ and back cancel;
on the other hand, it limits the temporal extension of the lattice universe
one can achieve without losing the baryonic Wilson loop signal to numerical
noise\footnote{Corresponding to this treatment of the baryonic potential,
also the mesonic string tension will be evaluated in section \ref{rotsb}
using Polyakov loop correlators.}.

\section{Numerical noise reduction}
To nevertheless reduce the numerical noise contaminating the mesonic
and baryonic Polyakov loop correlator measurements as far as possible,
the noise reduction technique introduced by L\"uscher and Weisz \cite{luwe}
was employed. Given an action which can be decomposed as
\begin{equation}
S[q] \equiv S[q^t , q^s ] = S^t [q^t ] + \sum_{i} S^i [q^t , q^s_i ] \ ,
\label{actdec}
\end{equation}
an observable $O$ which correspondingly factorizes as
\begin{equation}
O[q] \equiv O[q^t , q^s ] = \prod_{i} O^i [q^t , q^s_i ] \ ,
\label{obsdec}
\end{equation}
and a constraint $\delta $ which separates as
\begin{equation}
\delta [q] \equiv \delta [q^t , q^s ] = \delta^{t} [q^t ] 
\prod_{i} \delta^{i} [q^t , q^s_i ] \ ,
\label{consdec}
\end{equation}
the method of L\"uscher and Weisz corresponds to evaluating the expectation
value of $O$ over the set of variables $q$ subject to the constraint
$\delta $ using the action $S$ as
\begin{eqnarray}
\langle O \rangle_{\{ q ; \delta ; S \} } &=&
\frac{1}{Z} \int [Dq^t ] \, \delta^{t} [q^t ] \exp (-S^t [q^t ] )
\prod_{i} \int [Dq^s_i ] \delta^{i} [q^t , q^s_i ] 
O^i [q^t , q^s_i ] \exp (-S^i [q^t , q^s_i ] ) \nonumber \\
&=& \left\langle \prod_{i} 
\langle O^i \rangle_{\{ q^s_i ; \delta^{i} ; S^i \} } [q^t ]
\right\rangle_{\{ q ; \delta ; S \} }
\end{eqnarray}
In other words, the variables $q^t $ are kept fixed while the inner
expectation values are taken over their respective variables $q^s_i $
using the actions $S^i $ subject to the constraints $\delta^{i} $; in the
outer averaging over the full set of variables $q$ subject to the full
constraint $\delta $ using the full action $S$, the product of these inner
expectation values may be much easier to sample than the original
$O[q^t , q^s ]$ was. Note that, in practice, this method often can be
iterated in the sense that one may again select from the set of variables
$q^t $ a subset which is to be kept fixed while its complement
is being used for averaging; as long as the action, the constraint and
the quantity being averaged can be decomposed in analogy to
(\ref{actdec})-(\ref{consdec}), one can construct additional hierarchies
of averaging, where the quantities being averaged at each level only
depend on a subset of the variables the quantities at the next-lower
level depended on.

This method applies to the random vortex world-surface model if one
chooses as the variables $q^t $ all elementary squares which extend
into the time and one space direction, and as the variables $q^s_i $
the elementary squares at the $i$-th lattice time (which extend into
two spatial directions). The action (\ref{curvature}) is a sum over
terms associated with lattice links, each term coupling only
elementary squares attached to the corresponding link. The sum over
spatial links at the $i$-th lattice time thus yields the $S^i $ piece of
the action, whereas the sum over all temporal links yields the $S^t $ piece.
Similarly, the constraint of continuity of flux is to be satisfied
independently at each lattice link \cite{su3} by the elementary squares
attached to that link; it thus factorizes into terms coupling only the
variables $q^t $ and $q^s_i $ at fixed $i$, as well as a term
constraining only the variables $q^t $. Lastly, any Wilson loop can
be evaluated as a product over single plaquettes, which in turn each only
depend on the value of a single vortex elementary square according to
(\ref{plaq}); this can thus be trivially grouped into factors satisfying
the factorization (\ref{obsdec}). In accordance with the remarks on the
update procedure in section \ref{moddesc}, in this first level of
averaging, elementary updates are thus only performed on surfaces of
elementary lattice cubes which lie in a fixed time slice, i.e., which
extend into three spatial directions.

After this decomposition, in practice also a second averaging hierarchy
was employed; namely, the set of elementary squares $q^t $ extending into
the time and one space direction was decomposed into sets of squares
$q^t_{2i} $ connecting the $(2i-1)$-th lattice time with the $(2i)$-th
lattice time, with the remaining elementary squares making up the set
$\bar{q}^t $ to be kept fixed at this second level of averaging. With
this decomposition, the properties (\ref{actdec})-(\ref{consdec}) are
again satisfied; note that, for fixed $i$, the variables $q^t_{2i} $
only enter the product of inner expectation values
\[
\langle O^{2i-1} \rangle_{\{ q^s_{2i-1} ; \delta^{2i-1} ; S^{2i-1} \} }
\, \langle O^{2i} \rangle_{\{ q^s_{2i} ; \delta^{2i} ; S^{2i} \} } \ .
\]
Thus, in this second level of averaging, updates were performed on the
surfaces of all elementary lattice cubes except for the ones connecting
even lattice times $2i$ with the next {\em higher} odd lattice times
$2i+1$.

In practice, it turned out to be efficient to carry out the innermost
averaging using either 4000 or 8000 configurations; in the second-level
averaging, either 200 or 400 configurations were used. For the outermost
averaging, typically 20 to 60 configurations were enough to achieve a
sufficient level of accuracy.

\section{Angular dependence of the string tension}
\label{rotsb}
As already mentioned further above, the mesonic string tension was 
evaluated using Polyakov loop correlators, employing the noise reduction
techniques discussed in the previous section. In practice, $16^3 \times 4$
lattices were used, with the curvature coupling $c$ in (\ref{curvature})
set to the physical value $c=0.21$, cf.~\cite{su3}. The temporal extension
of four lattice spacings corresponds to a temperature of $T=0.45 T_c $
(where $T_c $ denotes the deconfinement temperature). In the case of
the static potential between sources separated along one of the lattice axes
studied in \cite{su3}, the string tension at this temperature deviated from
the extrapolation to zero temperature by less than 1~\%.

The mesonic geometries studied in the present work are listed in Table
\ref{table1}. The geometry labels defined there will be used below
to refer to the different cases.
\begin{table}[h]

\[
\begin{array}{|c|c|c|c|}
\hline
\mbox{Geometry} & \mbox{Relative spatial separation} & \mbox{Scale factors} &
\mbox{String tension in} \\
\mbox{label} & \mbox{of Polyakov loops} & n \ \mbox{used} &
\mbox{lattice units} \ \sigma a^2 \\
\hline\hline
M100 & n\cdot (a,0,0) & 1,2,3,4,5 & 0.7659 \pm 0.0004 \\ \hline
M110 & n\cdot (a,a,0) & 1,2,3 & 0.907 \pm 0.001 \\ \hline
M210 & n\cdot (2a,a,0) & 1,2 & 0.885^{+0.001}_{-0.027} \\ \hline
M310 & n\cdot (3a,a,0) & 1,2 & 0.83^{+0.015}_{-0.04} \\ \hline
M111 & n\cdot (a,a,a) & 1,2 & 1.02^{+0.002}_{-0.035} \\ \hline
M211 & n\cdot (2a,a,a) & 1,2 & 0.974^{+0.001}_{-0.03} \\ \hline
\end{array}
\]
\caption{Mesonic geometries and associated string tensions.}
\label{table1}
\end{table}
String tensions $\sigma $ were extracted by fitting the behavior of the
measured Polyakov loop correlators $P(nda)$ at different spatial distances
$nda$ separating the Polyakov loops by the ansatz
\begin{equation}
\frac{-\ln P(nda)}{4anda} = \sigma + \frac{\tau }{n}
\end{equation}
using the $n=1,2$ data, where $a$ denotes the lattice spacing\footnote{In
physical units, obtained \cite{su3} by equating the zero-temperature
string tension (along lattice axes) with $(440\, \mbox{MeV})^2 $, the
lattice spacing in the random vortex world-surface model is $0.39\, $ fm.}.
Of course, $4a$ is the temporal extension of the lattice, cf.~above; thus,
the denominator on the left hand side is simply
the minimal area spanned by the pair of Polyakov loops. The accuracy 
of this procedure was checked by also performing fits of the form
$\sigma + \tau /n +\rho /n^2 $ to the $n=1,2,3$ data in the 
$M100$ and $M110$ cases; this led to adjustments of the string tension
values by at most 3~\% (in both cases, downwards). These fits then did
also describe the $n=4,5$ data in the $M100$ case within their statistical
errors. In the error analysis of quantities derived from
string tensions extracted using only $n=1,2$ data, the aforementioned
3~\% systematical downwards uncertainty in these string tensions is
incorporated. The string tensions are listed in Table \ref{table1}.

\begin{figure}
\centerline{
\hspace{0.25cm}
\epsfig{file=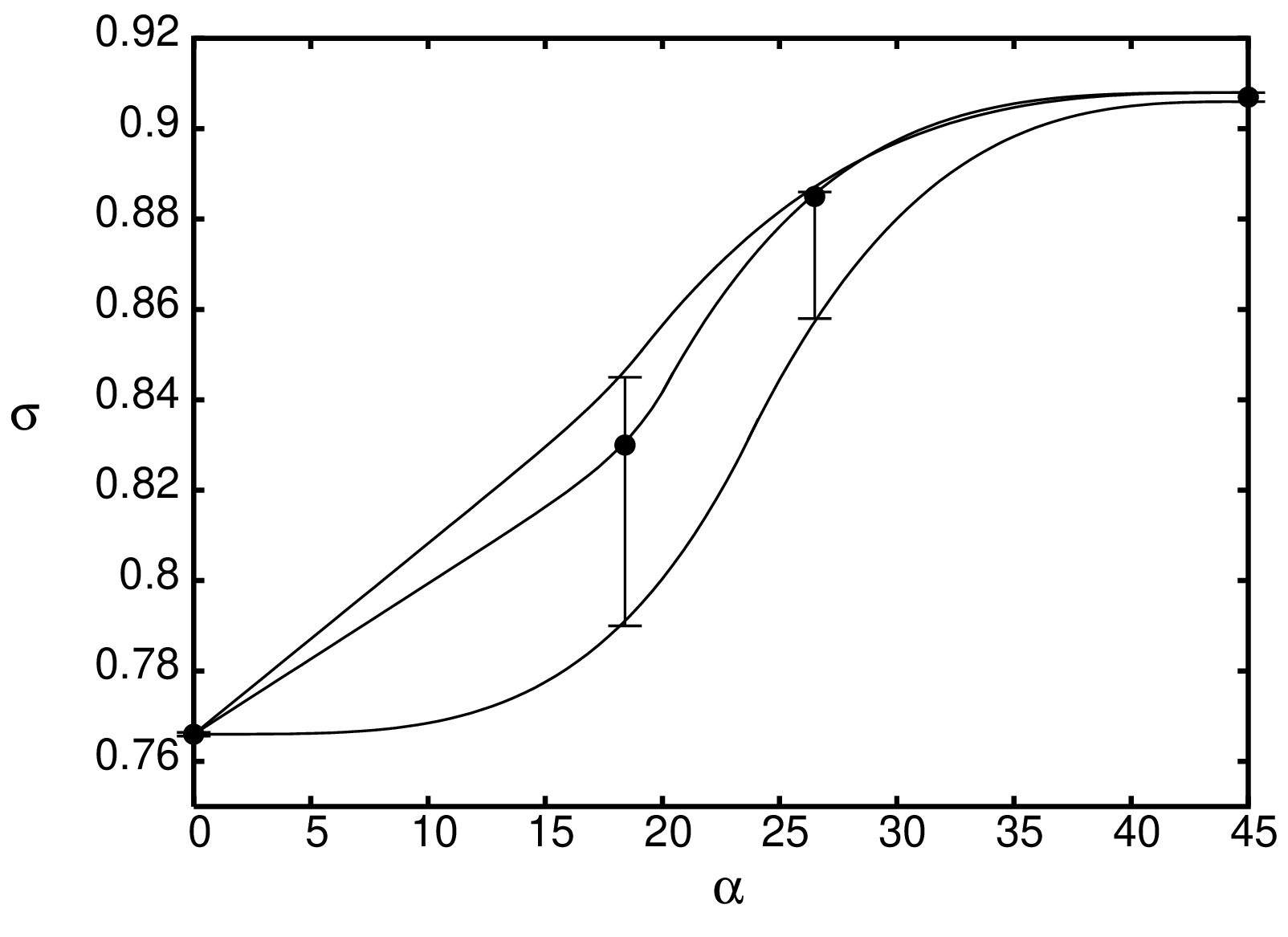,width=7.6cm}
\hspace{0.25cm}
\epsfig{file=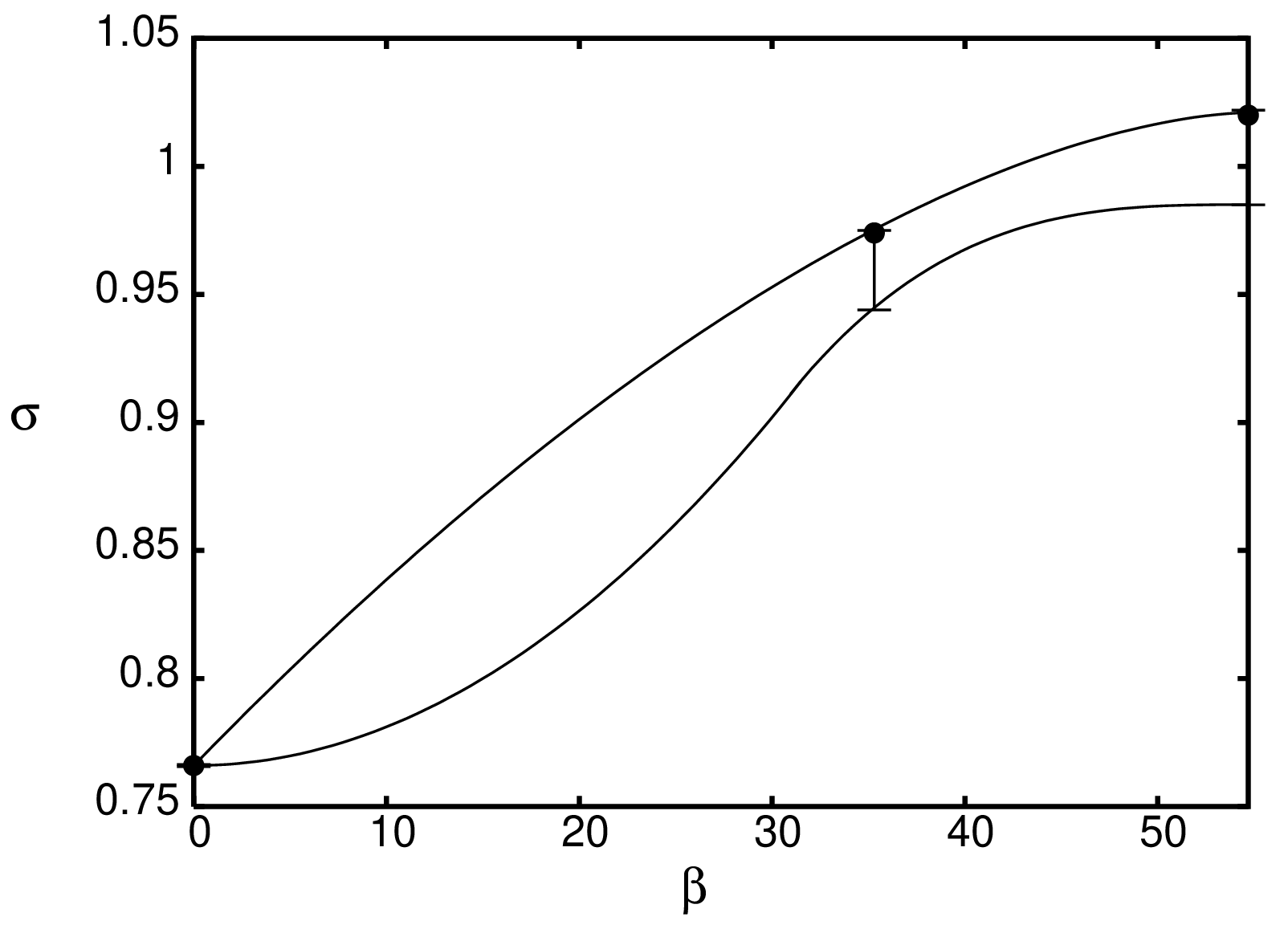,width=7.6cm}
}
\caption{\footnotesize
Angular dependence of the string tension $\sigma $. Left panel:
As a function of the angle $\alpha $ when the static sources are separated
along the direction $(\cos \alpha , \sin \alpha ,0)$; right panel: As a
function of the angle $\beta $ when the static sources are separated along
the direction
$((1/\sqrt{2} ) \sin \beta , (1/\sqrt{2} ) \sin \beta , \cos \beta )$.
Curves are fitted to the measured string tension values and to the extreme
values allowed by their error bars. In the case of $\sigma (\beta )$, the
fit to the measured values and the upper extreme coincide within the
resolution of the plot.}
\label{figab}
\end{figure}

Whereas $\Delta $ law predictions for the baryonic configurations studied
further below can be made purely on the basis of $\sigma_{M100} $ and
$\sigma_{M110} $, to arrive at Y law predictions, it is necessary to
interpolate continuous angular dependences of the mesonic string tension
from the discrete string tension data collected. To be specific, it is
necessary to combine
$\sigma_{M100} \, , \, \sigma_{M310} \, , \, \sigma_{M210} $ and
$\sigma_{M110} $ to interpolate the string tension for two sources
separated along the direction $(\cos \alpha , \sin \alpha ,0)$, and to
combine $\sigma_{M100} \, , \, \sigma_{M211} $ and $\sigma_{M111} $ to
interpolate the string tension for two sources separated along the direction
$((1/\sqrt{2} ) \sin \beta , (1/\sqrt{2} ) \sin \beta , \cos \beta )$.
Since the functional form of the $\alpha $- and $\beta $-dependences is
unknown, the quite conservative corridors for these dependences depicted
in Fig.~\ref{figab} were allowed for. Error bars for Y law predictions
below are based on the extremes allowed by these corridors.

\section{Baryonic geometries}
Baryonic potentials were measured by evaluating baryonic Polyakov loop
correlators as discussed in section \ref{obsdesc}. As in the mesonic
case, $16^3 \times 4$ lattices were used at the physical point $c=0.21$.
Table \ref{table2} lists the baryonic geometries studied\footnote{The
geometry labels $BL,BT,BS$ originate from the author's subjective picturing
of the respective configurations as ``L-shaped'', ``T-shaped'' and
``symmetric''.}.

\begin{table}[h]

\[
\begin{array}{|c|c|c|c|}
\hline
\mbox{Geometry} & \mbox{Static quark} & \mbox{Scale factors} &
\mbox{Baryonic tension in} \\
\mbox{label} & \mbox{positions} & n \ \mbox{used} &
\mbox{lattice units} \ V a^2 \\
\hline\hline
& (0,0,0) & & \\
BL & n\cdot (a,0,0) & 1,2,3 & 1.518 \pm 0.002 \\
& n\cdot (0,a,0) & & \\ \hline
& n\cdot (0,a,0) & & \\
BT & n\cdot (a,0,0) & 1,2 & 2.29^{+0.002}_{-0.07} \\
& n\cdot (-a,0,0) & & \\ \hline
& n\cdot (a,0,0) & & \\
BS & n\cdot (0,a,0) & 1,2 & 2.276^{+0.005}_{-0.073} \\
& n\cdot (0,0,a) & & \\ \hline
\end{array}
\]
\caption{Baryonic geometries and associated baryonic tensions.}
\label{table2}
\end{table}
Similar to the mesonic case, the leading long-range behavior of the
potential was extracted by fitting the measured baryonic Polyakov loop
correlators $P^B_n $ at the scale factors $n$ by the ansatz
\begin{equation}
\frac{-\ln P^B_n }{4ana} = V + \frac{U}{n}
\end{equation}
using the $n=1,2$ data. The baryonic tension $V$ can thus be directly
compared to $\Delta $ and Y law predictions obtained by adding bond lengths
(in units of $na$) weighted by the associated string tensions $\sigma $.
In the $BL$ case, fitting the $n=1,2,3$ data by $V+U/n+W/n^2 $ led to a
downwards adjustment of $V$ by 1.25~\%. The baryonic tensions $V$ are
listed in Table \ref{table2}. Motivated by the results of the fits in
the mesonic case, the baryonic tensions extracted using only $n=1,2$ data
were again also endowed with a 3~\% downwards systematic uncertainty.

$\Delta $ law predictions for these baryonic configurations are quite
straightforward, since the only relevant bond string tensions are
$\sigma_{M100} /2$ and $\sigma_{M110} /2$ (the reader is reminded that the
$\Delta $ model assumes half the mesonic string tensions to be associated
with the bonds). The predictions are
\begin{eqnarray}
V^{\Delta}_{BL} a^2 &=& 2 (\sigma_{M100} a^2 /2) +
\sqrt{2} (\sigma_{M110} a^2 /2) = 1.407 \pm 0.001 \\
V^{\Delta}_{BT} a^2 &=& 2\sqrt{2} (\sigma_{M110} a^2 /2) +
2 (\sigma_{M100} a^2 /2) = 2.0485 \pm 0.0015 \\
V^{\Delta}_{BS} a^2 &=& 3\sqrt{2} (\sigma_{M110} a^2 /2) = 1.924 \pm 0.002
\end{eqnarray}
On the other hand, to arrive at Y law predictions, one must perform a
potential energy minimization as a function of the bond junction position;
the interplay of varying bond lengths and associated string tensions
may favor a different bond junction than the one which would be favorable
if the string tension were perfectly isotropic. Since the string tension
is enhanced along diagonal directions, the optimal bond geometry will be
more aligned with the lattice axes than in the isotropic case. Exploring
bond junction positions in the spatial 1-2-plane for the $BL$ and $BT$
configurations and minimizing the associated potential energy for the range
of angular dependences $\sigma (\alpha )$ depicted in Fig.~\ref{figab}, one
arrives at the Y law predictions
\begin{eqnarray}
V^{Y}_{BL} a^2 &=& 1.523 \pm 0.008 \\
V^{Y}_{BT} a^2 &=& 2.25^{+0.02}_{-0.08} \ .
\end{eqnarray}
Specifically, the expectation is determined by using for $\sigma (\alpha )$
the fit to the measured string tension values displayed in Fig.~\ref{figab},
whereas the error bars are determined by using the most extreme
$\sigma (\alpha )$ dependences allowed for by Fig.~\ref{figab}.

Treatment of the $BS$ geometry is more complicated. If one assumes the
$60^o $ discrete rotational symmetry of the configuration to be unbroken,
then the bond junction position is located in the $(1,1,1)$ direction
viewed from the origin and one can straightforwardly use the range of
angular dependences $\sigma (\beta )$ allowed for by Fig.~\ref{figab}
to analyze this restricted case, in complete analogy to the treatment of
the $BL$ and $BT$ configurations above. Searching for the minimal potential
energy under this assumption to arrive at a first estimate regarding the
Y law prediction for the $BS$ geometry, one arrives at
\begin{equation}
\left. V^{Y}_{BS} a^2 \right|_{(1,1,1)} = 2.271^{+0.002}_{-0.18} \ .
\label{restbs}
\end{equation}
On the other hand, there is no guarantee that the bond junction will
indeed be located in the $(1,1,1)$ direction viewed from the origin.
It is therefore necessary to explore the whole three-dimensional range
of possible bond junction positions; however, the associated potential
energies in general can only be given if the general angular dependence
of the string tension is supplied. To arrive at an estimate for this
dependence, the particular cases $\sigma (\alpha )$ and
$\sigma (\beta )$ displayed in Fig.~\ref{figab}, which essentially give
the string tension at $0^o $ and $45^o $ longitude on a sphere (if one
additionally uses $\sigma (\beta =90^o ) = \sigma (\alpha =45^o )$), were
used to construct an interpolation of the string tension for other
longitudes $\varphi $. To be specific, the form $\sigma (\varphi ) =
(1-\sin(2\varphi )) \sigma (\varphi =0^o ) +
\sin(2\varphi ) \sigma (\varphi =45^o )$ was assumed, which also
reasonably well fits the additional information that the dependence on
the longitude $\varphi $ at the ``equator'', i.e., connecting
$\sigma (\alpha = 90^o )$ with $\sigma (\beta = 90^o )$, must again take
the same functional form as $\sigma (\alpha )$. This interpolation was
carried out both for the fits to the measured string tension values in
Fig.~\ref{figab} as well as for the upper and lower extremes, in order to
again arrive at an error estimate for the Y law generated. On this
basis, exploring the whole three-dimensional range of possible bond
junction positions indeed yields a slightly more favorable asymmetric
bond junction position and an associated Y law prediction of
\begin{equation}
V^{Y}_{BS} a^2 = 2.25^{+0.02}_{-0.16}
\end{equation}
which only deviates little from (\ref{restbs}); in analogy to above, the
error bars are based on minimizing the potential energy using the
$\varphi $-interpolations of the extreme $\sigma (\alpha )$ and
$\sigma (\beta )$ dependences allowed for by Fig.~\ref{figab}. Thus, the
systematic error associated with restricting the range of bond junction
positions in arriving at (\ref{restbs}) appears to be comparatively minor.

\section{Conclusions}
Carefully taking into account the anisotropy in the string tension
introduced into the random vortex world-surface model by the hypercubic
treatment of the vortex surfaces, one obtains an unambiguous characterization
of the behavior of the baryonic potential. Assembling the measured potentials
and the corresponding $\Delta $ and Y law predictions from the previous 
section, cf.~Table~\ref{table3},
\begin{table}[h]

\[
\begin{array}{|c|c|c|c|}
\hline
\mbox{Geometry} & Va^2 & V^Y a^2 & V^{\Delta } a^2 \\
\hline\hline
BL & 1.518 \pm 0.002 & 1.523 \pm 0.008 & 1.407 \pm 0.001 \\ \hline
BT & 2.29^{+0.002}_{-0.07} & 2.25^{+0.02}_{-0.08} & 2.0485 \pm 0.0015 \\
\hline
BS & 2.276^{+0.005}_{-0.073} & 2.25^{+0.02}_{-0.16} & 1.924 \pm 0.002 \\
\hline
\end{array}
\]
\caption{Comparison of measured baryonic tensions with $\Delta $ and Y law
predictions.}
\label{table3}
\end{table}
it is evident that the $\Delta $ law does not agree with the measured
baryonic potential, whereas the Y law yields rather accurate predictions
for it. It should be noted that the Y law behavior is seen quite clearly
already at moderate distances in the vortex model, contrary to full lattice
Yang-Mills theory \cite{phil02}, where it is necessary to consider rather
large distances. In the full theory, presumably the short-distance
perturbative $\Delta $ behavior masks the long-distance Y law and pushes
its onset to larger separations. In the vortex model, by contrast,
such perturbative effects are truncated and the Y law induced by the
vortices clearly dominates the behavior of the baryonic potential.

\section*{Acknowledgments}
The author is grateful to M.~Quandt and H.~Reinhardt for fruitful
discussions on the $SU(3)$ random vortex world-surface model, and to
P.~de~Forcrand for enlightening correspondence on aspects of the
baryonic potential. Furthermore, science+computing ag, T\"ubingen,
is acknowledged for providing computational resources. This work
was supported in part by the U.S.~DOE under grant number DE-FG03-95ER40965.


\begin{thebibliography}{99}
\bibitem{vormod} M.~Engelhardt and H.~Reinhardt,
Nucl. Phys. {\bf B585} (2000) 591.
\bibitem{topol} M.~Engelhardt, Nucl. Phys. {\bf B585} (2000) 614.
\bibitem{csb} M.~Engelhardt, Nucl. Phys. {\bf B638} (2002) 81.
\bibitem{su3} M.~Engelhardt, M.~Quandt and H.~Reinhardt, Nucl. Phys. 
{\bf B685} (2004) 227.
\bibitem{deb97} L.~Del Debbio, M.~Faber, J.~Greensite and
\v{S}.~Olejn{\'\i}k, Phys. Rev. {\bf D 55} (1997) 2298.
\bibitem{giedt} L.~Del Debbio, M.~Faber, J.~Giedt, J.~Greensite and
\v{S}.~Olejn{\'\i}k, Phys. Rev. {\bf D 58} (1998) 094501.
\bibitem{fabdo3} M.~Faber, J.~Greensite and \v{S}.~Olejn{\'\i}k,
Phys. Lett. {\bf B474} (2000) 177.
\bibitem{greensrev} J.~Greensite, Prog. Part. Nucl. Phys. {\bf 51} (2003) 1.
\bibitem{bertop} R.~Bertle, M.~Engelhardt and M.~Faber,
Phys. Rev. {\bf D 64} (2001) 074504.
\bibitem{temp} K.~Langfeld, O.~Tennert, M.~Engelhardt and H.~Reinhardt,
Phys. Lett. {\bf B452} (1999) 301.
\bibitem{tlang} M.~Engelhardt, K.~Langfeld, H.~Reinhardt and O.~Tennert,
Phys. Rev. {\bf D 61} (2000) 054504.
\bibitem{forc1} P.~de~Forcrand and M.~D'Elia,
Phys. Rev. Lett. {\bf 82} (1999) 4582.
\bibitem{forc2} C.~Alexandrou, M.~D'Elia and P.~de~Forcrand,
Nucl. Phys. Proc. Suppl. {\bf 83} (2000) 437.
\bibitem{pepe} P.~de~Forcrand and M.~Pepe, Nucl. Phys. {\bf B598} (2001) 557.
\bibitem{corndlaw} J.~M.~Cornwall, Phys. Rev. {\bf D 54} (1996) 6527.
\bibitem{cornylaw} J.~M.~Cornwall, Phys. Rev. {\bf D 69} (2004) 065013.
\bibitem{takahashi} T.~T.~Takahashi, H.~Matsufuru, Y.~Nemoto and H.~Suganuma,
Phys. Rev. Lett. {\bf 86} (2001) 18;\\
T.~T.~Takahashi, H.~Suganuma, Y.~Nemoto and H.~Matsufuru,
Phys. Rev. {\bf D 65} (2002) 114509.
\bibitem{phil01} C.~Alexandrou, P.~de~Forcrand and A.~Tsapalis,
Phys. Rev. {\bf D 65} (2002) 054503.
\bibitem{phil02} C.~Alexandrou, P.~de~Forcrand and O.~Jahn,
Nucl. Phys. Proc. Suppl. {\bf 119} (2003) 667.
\bibitem{philoli} O.~Jahn and P.~de~Forcrand,
Nucl. Phys. Proc. Suppl. {\bf 129} (2004) 700.
\bibitem{cherkom} M.~N.~Chernodub and D.~A.~Komarov,
JETP Lett. {\bf 68} (1998) 117.
\bibitem{ichie} V.~G.~Bornyakov, H.~Ichie, Y.~Mori, D.~Pleiter,
M.~I.~Polikarpov, G.~Schierholz, T.~Streuer, H.~St\"uben and T.~Suzuki,
hep-lat/0401026.
\bibitem{stochv} D.~S.~Kuzmenko and Yu.~A.~Simonov, Phys. Lett. {\bf B494}
(2000) 81.
\bibitem{cont} M.~Engelhardt and H.~Reinhardt,
Nucl. Phys. {\bf B567} (2000) 249.
\bibitem{spag} H.~B.~Nielsen and P.~Olesen, Nucl. Phys. {\bf B160}
(1979) 380; \\
J.~Ambj{\o}rn and P.~Olesen, Nucl. Phys. {\bf B170}
[FS1] (1980) 60; \\
J.~Ambj{\o}rn and P.~Olesen, Nucl. Phys. {\bf B170}
[FS1] (1980) 265.
\bibitem{kogut} J.~B.~Kogut, Rev. Mod. Phys. {\bf 55} (1983) 775.
\bibitem{luwe} M.~L\"uscher and P.~Weisz, JHEP {\bf 0109} (2001) 010.
\end{thebibliography}
\end{document}